\documentclass[a4paper,11pt]{article}
\pdfoutput=1 

\usepackage{jcappub} 

\usepackage[T1]{fontenc} 

\title{\boldmath Symmetrically Tuned Large-Volume Conic Shell-Cavities for Axion Searches}

\author[a,b]{Chao-Lin Kuo}

\affiliation[a]{Kavli Institute of Particle Astrophysics and Cosmology \& Physics Department,
Stanford University, Stanford, CA 94305, USA}
\affiliation[b]{SLAC National Accelerator Laboratory, Menlo Park, CA 94025, USA}

\emailAdd{clkuo@stanford.edu}

\abstract{In an earlier paper \cite{Kuo_2020}, a new class of thin-shell cavities were proposed to evade the steep frequency scaling of conventional axion haloscopes.  In this follow-up work, we see that a generalized conic geometry enables robust frequency-tuning for these large-volume cm-wave cavities.  The frequency-defining dimension of a conic shell-cavity changes symmetrically and uniformly during tuning, maintaining a high axion coupling efficiency (the ``form factor'') to an external solenoid field.  It is further shown that such tunable geometry is not restricted to circular cones.  A general prescription for arbitrary volume-filling conic shell-cavities is developed and direct solutions are obtained for the created numerical models.  The largest of the realized designs is a meandering ``brain'' cavity that is tunable over a frequency range of $20\%$.  The scan rate of this cavity is three orders of magnitude larger than that of a scaled cylindrical cavity used in the current generation experiments. The prospects for such a large improvement in the scan rate should motivate R\& D efforts in fabrication and other implementation techniques.  If these engineering challenges can be met, cavity-based axion haloscopes can stay competitive at frequencies higher than a few GHz. We propose an experimental configuration at $20$ GHz ($\sim 80\;\mu$eV) using an array of brain cavities and compare it with other proposals for similar frequencies.}

\begin{document}
\maketitle

\flushbottom

\section{Introduction}

The scan rate of a cavity-based axion dark matter search \cite{pq,weinberg,wilczek,sikivie1,abbott83,dine83,preskill83} is proportional to the square of the volume. 
The steep frequency scaling of the cavity volume ($V\propto\nu^{-3}$) leads to significant experimental challenges for frequencies higher than a few GHz.  This frequency (or mass $m_a$) range corresponds to the post-inflationary scenario \cite{hertzberg,marsh,graham2,Takahashi_18,Nakagawa_2020,diluzio_20}, which generally predicts an axion mass of $20-120 \;\mu$eV ($5-30$ GHz) if QCD axion is to account for {\em all} of the dark matter \cite{Borsanyi_16,Klaer_2017,Kawasawi_18,Hindmarsh_20,Buschmann_20}.  Despite some modeling uncertainties, this classical axion window contains no fudge factors because the only free parameter is already fixed by the observed dark matter density.  Interestingly, this scenario implies a high inflationary energy scale $E_I$ and a level of gravitational wave background potentially detectable in the CMB (cosmic microwave background) as $B$-mode polarization.  Conversely in the pre-inflationary (also known as the ``stochastic'') axion scenario, the observed dark matter density can only be produced with a very low inflationary energy scale that leads to an undetectable primordial $B$-mode polarization.  
On the customary $E_I-m_a$ plane for QCD axions \cite{graham2,diluzio_20}, the post-inflationary scenario is sandwiched by the BICEP2/Keck results on one end \cite{bk15} and Planck's isocurvature constraints on the other \cite{Planck_inf}. 
A vigorous program aiming to detect or rule out QCD axion in the cm-wave frequencies is therefore highly synergistic with the intense global effort to measure the inflationary energy scale via the primordial $B$-modes \cite{BA_18,so_19,s4science,litebird}. 

In an earlier paper \cite{Kuo_2020} (Paper I), a new class of thin-shell cavities were proposed to search for QCD axions in the cm-wave frequency range. 
Like a conventional cylindrical cavity, a thin-shell cavity has a high quality factor $Q$ and a singly polarized TM$_{010}$ mode that facilitates efficient axion-photon conversion in the presence of a uniform external magnetic field.  The most notable new feature is that the active volume scales as $\propto \nu^{-1}$, or even milder if a volume-filling ``brain'' variety can be realized.  These cavities evade the  $\nu^{-3}$ scaling by decoupling the resonant frequency from the volume-defining ({\em transverse}) dimensions \cite{Kuo_2020}.  This improvement makes the thin-shell geometry overwhelmingly advantageous in the cm-wave frequencies.  Prior to the publication of Paper I, a conventional wisdom was that the dimensions of a cavity haloscope cannot be larger than the Compton wavelength.

\begin{figure}[t]
\centering 
\includegraphics[width=6in]{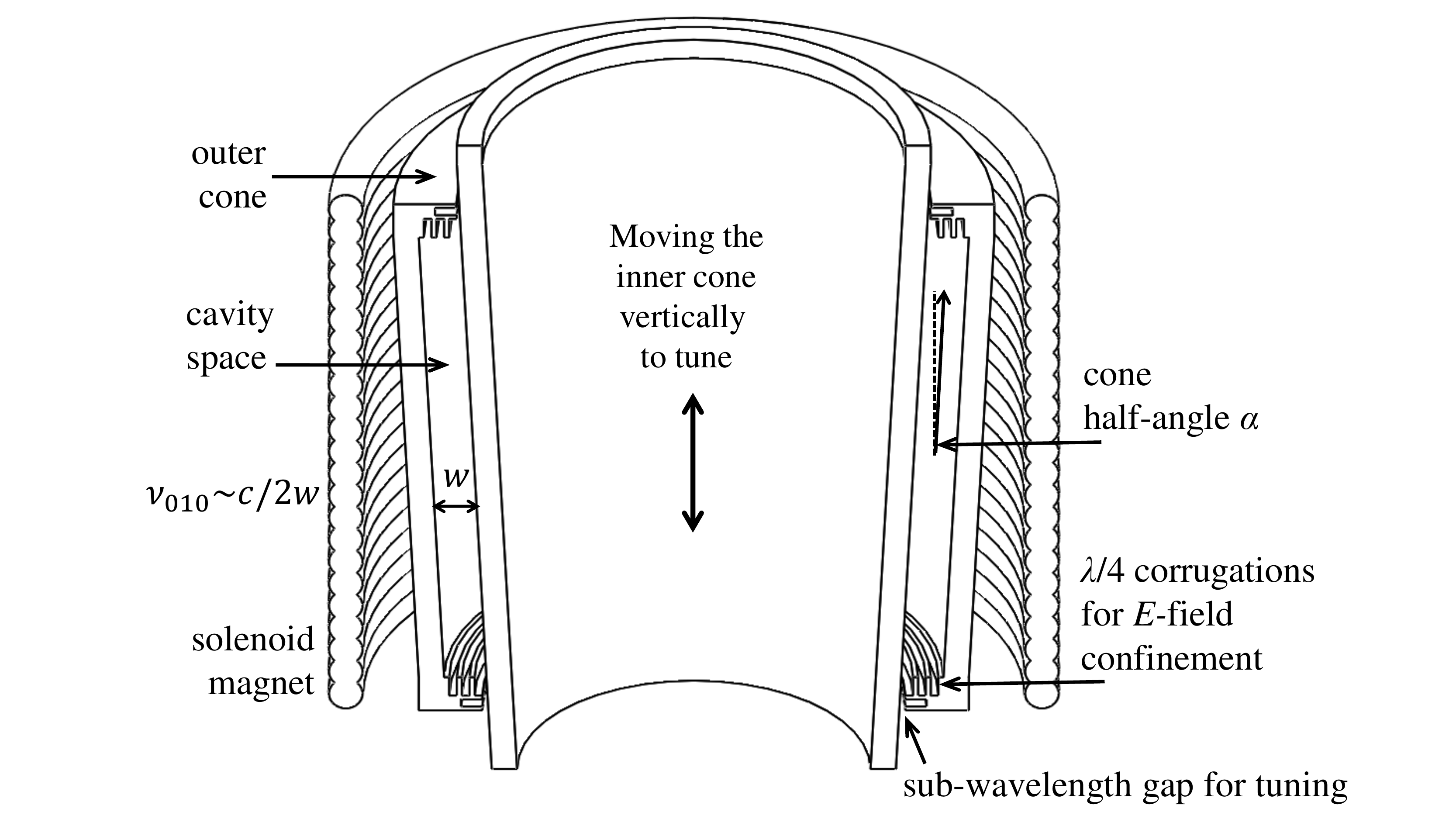}
\caption{\label{fig:cupholder} The schematic diagram of a conic shell cavity for axion searches. The cavity space is formed between two nested cones. Frequency-tuning is achieved with vertical motions of the inner cone.  As discussed in Paper I \cite{Kuo_2020}, $\lambda/4$-corrugations are used to confine the mode in the presence of a inter-cone gap that allows $20\%$ frequency tuning.  During the full range of tuning, the distance $w$ between the two cones remains uniform, avoiding asymmetry-induced volume losses.  In this figure, the cone half-angle $\alpha$ is small ($<3^\circ$), facilitating strong ${\bf E}\cdot {\bf B}$ inverse Primakoff coupling to the external solenoid $B$-field in the $z$ direction.  Detailed numerical results of this cavity (labeled ``Cup Holder'') and other designs with even larger volume are discussed in Section 3.}
\end{figure}

In this paper (Paper II), the thin-shell design is advanced further.  The most significant conceptual development is a conic geometry that enables a robust tuning scheme (Figure \ref{fig:cupholder}).  The frequency-defining dimension $w$ of the cavity changes symmetrically during tuning, avoiding disproportional volume losses caused by lateral tuning.  It is further shown that a tunable conic cavity can be derived from a nearly arbitrary closed curve.  Several varieties of such generalized conic designs have been created and modeled.  Explicit numerical solutions have been obtained using FEA (Finite Element Analysis) methods, including those for a volume-filling brain cavity that can improve the scan rate by three orders of magnitude over conventional designs.  The quality factor $Q$ and form factor $C_{010}$ confirm the scaling arguments presented in Paper I. 
It is found that the corrugations proposed in Paper I to suppress boundary $E$ fields also suppress spurious TEM modes. As a result, the loss in frequency range due to mode crossing is modest. 

In Section 2, a prescription is given to create a {\em generalized} conic shell-cavity.  Section 3 compares numerical results for a few conic thin-shell cavities with reference cylindrical cavities. Section 4 outlines a strawperson configuration at $20$ GHz and concludes the paper with comparisons of the shell-cavity proposal to two other techniques suggested for the similar frequency range: dielectric haloscopes \cite{madmax_17} and plasma haloscopes \cite{plasma}. 

\section{Constructing a {\em Tunable} Conic Shell-Cavity}


In a thin-shell cavity, the eigenmode used for axion searches can be seen as a $z-$polarized quasi-plane wave bouncing back and forth between the inner and outer walls, setting up a standing wave.  Such an eigenmode is generally called TM$_{010}$-like mode in the literature, or simply as TM$_{010}$.  As pointed out in \cite{Kuo_2020}, the two transverse dimensions can be exploited to increase the cavity volume and ultimately the scan rate, limited only by implementation practicalities.  

\begin{figure}[t]
\centering 

\includegraphics[width=6.in]{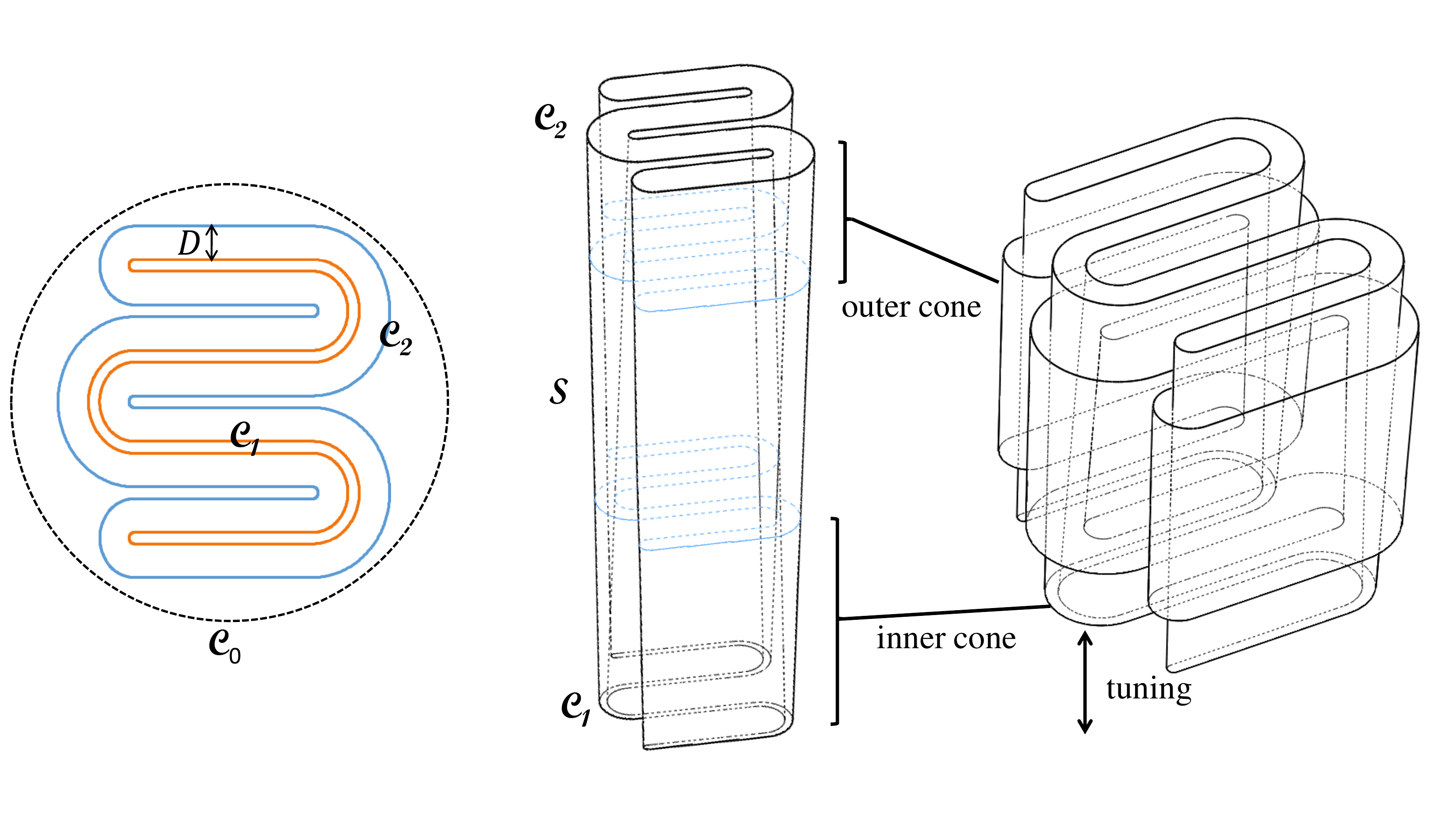}
\caption{\label{fig:W} The circular cone depicted in Figure \ref{fig:cupholder} can be generalized to have an even larger volume.  This figure illustrates the construction of a tunable volume-filling conic shell cavity.  Two closed curves ${\cal C}_1$ and ${\cal C}_2$ separated by $D$ ({\em Left}) defines a loft surface ${\cal S}$ ({\em Middle}).  The inner and outer cones of a tunable shell cavity ({\em Right}) are taken from the lower and upper sections of ${\cal S}$.  As in the case of circular conic shells, the tuning is achieved with vertical motions of the inner cone which keep the distance between the two cones uniform by design. Small ($<1\%$) adjustments of $D$ along ${\cal C}_2$ are necessary to correct for the effects of turn-radii. Numerical results of this cavity (labeled ``W-Meander'') are presented in Section 3.  The axion scan rate  of this cavity is three orders of magnitude larger than a conventional cylindrical cavity.  
}
\end{figure}

As originally conceived, a thin-shell cavity would have vertical inner and outer high-conductivity metallic walls. Paper I proposes to tune the resonant frequency of the TM$_{010}$ mode by laterally displacing the inner assembly similar to conventional axion cavities \cite{hagmann90, lyapustin}. There are two major disadvantages to this approach. First of all, the broken azimuthal symmetry leads to mode localization and a corresponding reduction in form factor, which measures the coupling efficiency to an external magnetic field. This leads to significant losses in the scan speed of thin-shell cavities. Secondly, this scheme is difficult to implement in a brain cavity which maximizes gains in scan rates and could make cavities competitive in the crucial cm-wave frequency range. 

A key modification to the original concept avoids both issues. The new scheme is easily understood in the simplest case of circular geometries.  Instead of two nested straight cylinders, the new geometry features two {\em cones}, defining the cavity space in between (Figure \ref{fig:cupholder}). 
As the inner cone moves vertically, the distance $w$ between the two cones changes {\em uniformly} throughout the cavity space, leading to predictable linear changes in the resonant frequency.  In retrospect, this ``coffee cup $\&$ holder'' geometry seems obvious. Nevertheless, to our knowledge no similar cavities have been proposed as tunable resonators, much less for axion searches. 
If the cone half-angle $\alpha$ is small, the geometry is only slightly different from a cylinder. One can conjecture that an eigenmode with high $Q$ still resembles TM$_{010}$ and the loss of the form factor would be on the order of $\cos(\alpha)$ and therefore insignificant.  This indeed has been directly verified with numerical calculations detailed in Section 2. 

The cavity in Figure \ref{fig:cupholder} obviously does not effectively use all of the magnetized volume.  In Paper I, it was hypothesized that a meandering volume-filling shell (dubbed the ``brain'' cavity) can also support a similar singly polarized eigenmode.  But no demonstrations of such have been given in that paper.  Driven by the quest for sensitivity, an explicit design of a symmetrically tuned {\em brain} cavity is given below. 

The mathematical construction of a generalized conic shell-cavity is illustrated in Figure \ref{fig:W}. The starting point is a closed plane curve ${\cal C}_1$ (Figure \ref{fig:W}, {\em Left}), and a second curve ${\cal C}_2$ as the collection of points at a fixed distance $D$ from ${\cal C}_1$. The curve ${\cal C}_2$ needs to stay within the (circular) cross section ${\cal C}_0$ of the solenoid $B$-field. Other then that, ${\cal C}_1$ can be nearly arbitrary.  For clarity, the curve ${\cal C}_1$ chosen in Figure \ref{fig:W} only meanders modestly into the shape of the letter W. To maximize the volume (and therefore the scan rate) of the eventual axion cavity, ${\cal C}_1$ can be a {\em closed} version of space-filling curves {\em \`a la} G. Peano \cite{Peano} and D. Hilbert \cite{Hilbert}. If fabrication challenges can be met, this could be a game-changer for higher frequencies ($>10$ GHz) and makes cavities stay competitive.

The next step is to vertically displace ${\cal C}_2$ by a distance of $D/\tan{\alpha}$, where $\alpha$ is the cone half-angle.  On each plane between ${\cal C}_1$ and ${\cal C}_2$, define a curve $\cal C$ to be the linear combination of ${\cal C}_1$ and ${\cal C}_2$. A cone section ${\cal S}$ is the loft surface swept by ${\cal C}$ as it linearly morphs from ${\cal C}_1$ to ${\cal C}_2$ over this vertical distance (Figure \ref{fig:W}, {\em Middle}). A large-volume shell cavity can then be formed by taking an upper (wider) section of ${\cal S}$ to be the outer cavity cone and a lower (narrower) section to be the inner cavity cone (Figure \ref{fig:W}, {\em Right}).  It is easy to see that the cavity defined between the two cones maintains a uniform spacing that is tunable by moving the inner cone vertically. 

It should be noted that $\cal S$ does not extrapolate to a point. Therefore, it is not truly part of a Generalized Cone defined in mathematics textbooks ({\em e.g.} \cite{weisstein} and references therein).  On the other hand, $\cal S$ is by construction a {\em ruled} surface. This means that through any given point on $\cal S$, there is a direction in which the surface follows a straight line \cite{weisstein2}. This may simplify the machining and metrology of the cavities.  
A slight complication is that although the inter-wall distance remains uniform throughout the cavity, the local resonant frequency varies slightly with the radius of curvature.  To maintain a high quality factor and a large form factor, this perturbation must be corrected by varying $D$ along ${\cal C}_2$.  In Paper I, an analytic formula based on asymptotic expressions of Bessel functions was given.  This can also be corrected numerically, which is what has been done to produce the results in Section 3 of this paper. 

\section{Numerical Results}

In this section, 3D numerical results for conic thin-shell cavities are presented. All calculations have been done with COMSOL-RF on an 18-core workstation with 96 GB of memory.  

Results from three types of conic thin-shell cavities are compared with two reference designs: (1) a scaled HAYSTAC cavity and (2) a cylindrical thin-shell cavity as proposed in Paper I.  Cavity models are created using the commercial software SolidWorks following the procedure described in the previous section.  Eigenmodes are then solved for these initial models.  As mentioned earlier, varying turn-radii split the mode into several modes that are adjacent in frequency.  The modes with high axion form factors $(\int E_z dV)^2/(V\int E^2 dV)$ are identified and the solid models are adjusted by locally changing $D$ (Figure \ref{fig:W}) to coalesce these modes into a single resonance.  Empirically, satisfactory results can be achieved with fewer than $3-5$ iterations for the ``W-Meander''.  The results in this paper were all obtained with adjustments based on eigenmode distributions at the center frequency. It is certainly possible to obtain even better performance with global optimization across the entire range of frequency.

\begin{figure}[b]
\centering 
\includegraphics[width=6in]{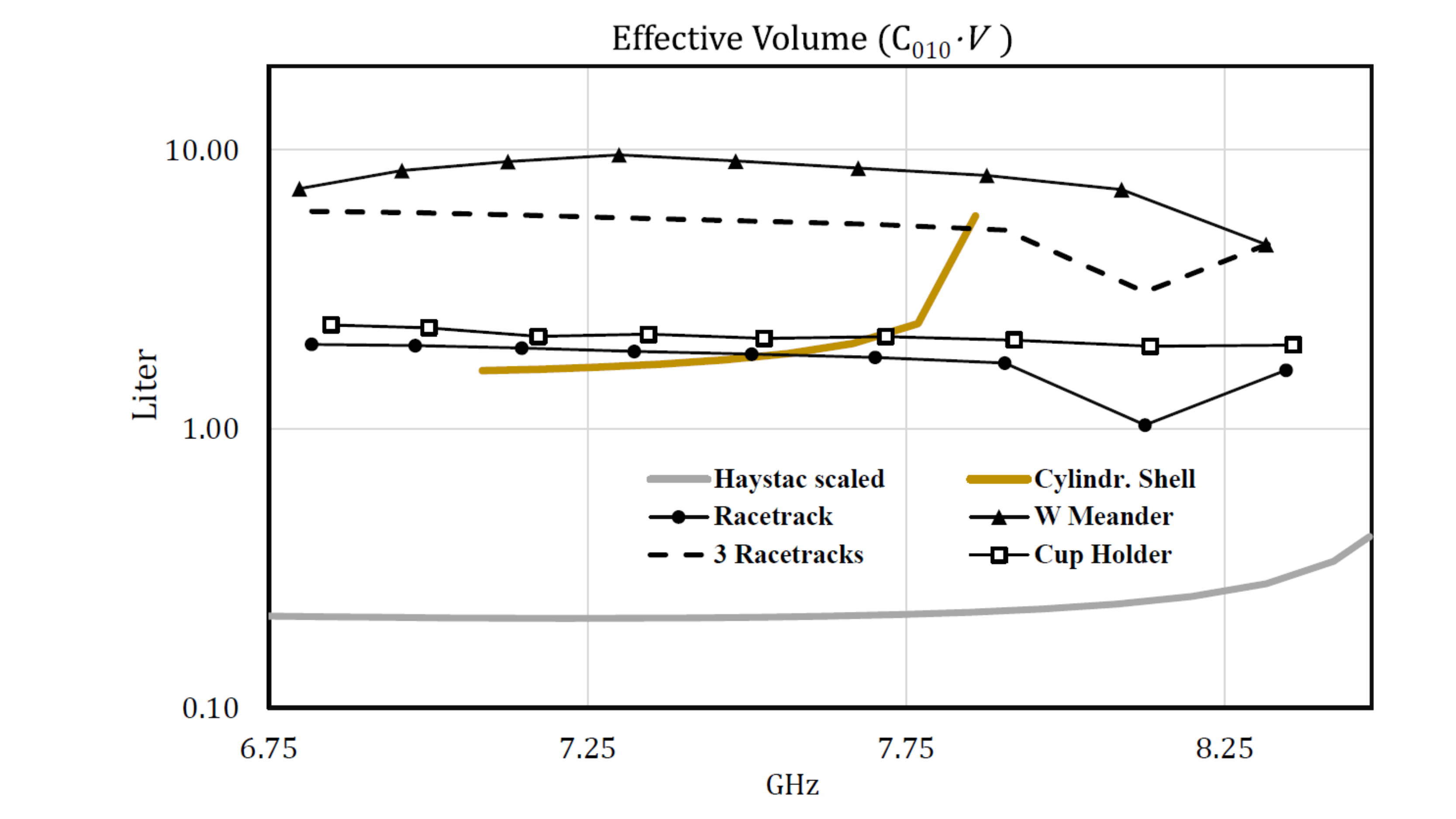}
\caption{\label{fig:EV} The scan rate of an axion haloscope is proportional to the square of the effective mode volume, $C_{010}V$.  This plot uses this figure of merit to compare three conic shell designs (W-Meander, Racetrack, and Cup Holder) against two reference designs (Scaled HAYSTAC and Cylindr. Shell).  }
\end{figure}

\subsection{The Reference Designs}

As a reference for comparison, a conventional cylindrical cavity based on the HAYSTAC \cite{brubaker} geometry was modeled.  The center frequency is placed at $7.5$ GHz as with other designs studied in this paper.  FEA confirms known basic results for this cavity: the existence of a TM$_{010}$-like mode with a form factor $C_{010}$ of $0.81-0.42$ over the tuning range. The effective mode volume, defined to be the product of the cavity volume and the form factor, is plotted as a function of frequency in Figure \ref{fig:EV} for each cavity. The scan rate of an axion haloscope is proportional to the square of this figure-of-merit. 

A second reference design is the baseline non-conic large-volume cavity, labeled as ``Cylindr. Shell'' in Figure \ref{fig:EV}. This is obtained by increasing the radius of the tuning rod of the ``scaled HAYSTAC'' in sync with the radius of the outer cylinder.  Like the scaled HAYSTAC, tuning is achieved with lateral displacements.  It is found that the form factor $C_{010}$ drops quickly from $0.81$ to $0.23$ when the inner cylinder is displaced off-center for tuning.  Furthermore, FEA fails to find the TM$_{010}$ mode when the resonance is tuned more than $10\%$ lower.  It is unclear whether this is fundamental or numerical.  The gain from the larger volume is still substantial but somewhat diminished by these side-effects.  

\subsection{W-Meander and Racetracks}

A major result of this paper is that the volume-filling design created in Figure \ref{fig:W} produces  a vastly larger effective volume than that of a conventional cavity over $>20\%$ tunable frequency range (Figure \ref{fig:EV}, labele ``W-Meander''). In the realized numerical model, the distance $D$ is set to be $34$ mm, and the cone half-angle $\alpha=\tan^{-1}{0.025}=1.43^\circ$. The inter-cone distance varies between $18$ mm and $22$ mm as the inner cone moves vertically.  Across the entire tuning range, the W-Meander cavity can provide $2000\times$ the scan rate of a scaled HAYSTAC. The efficiency decreases only modestly at the upper and lower ends of the range (see Figure \ref{fig:EV}).

\begin{figure}[b]
\centering 
\includegraphics[width=6in]{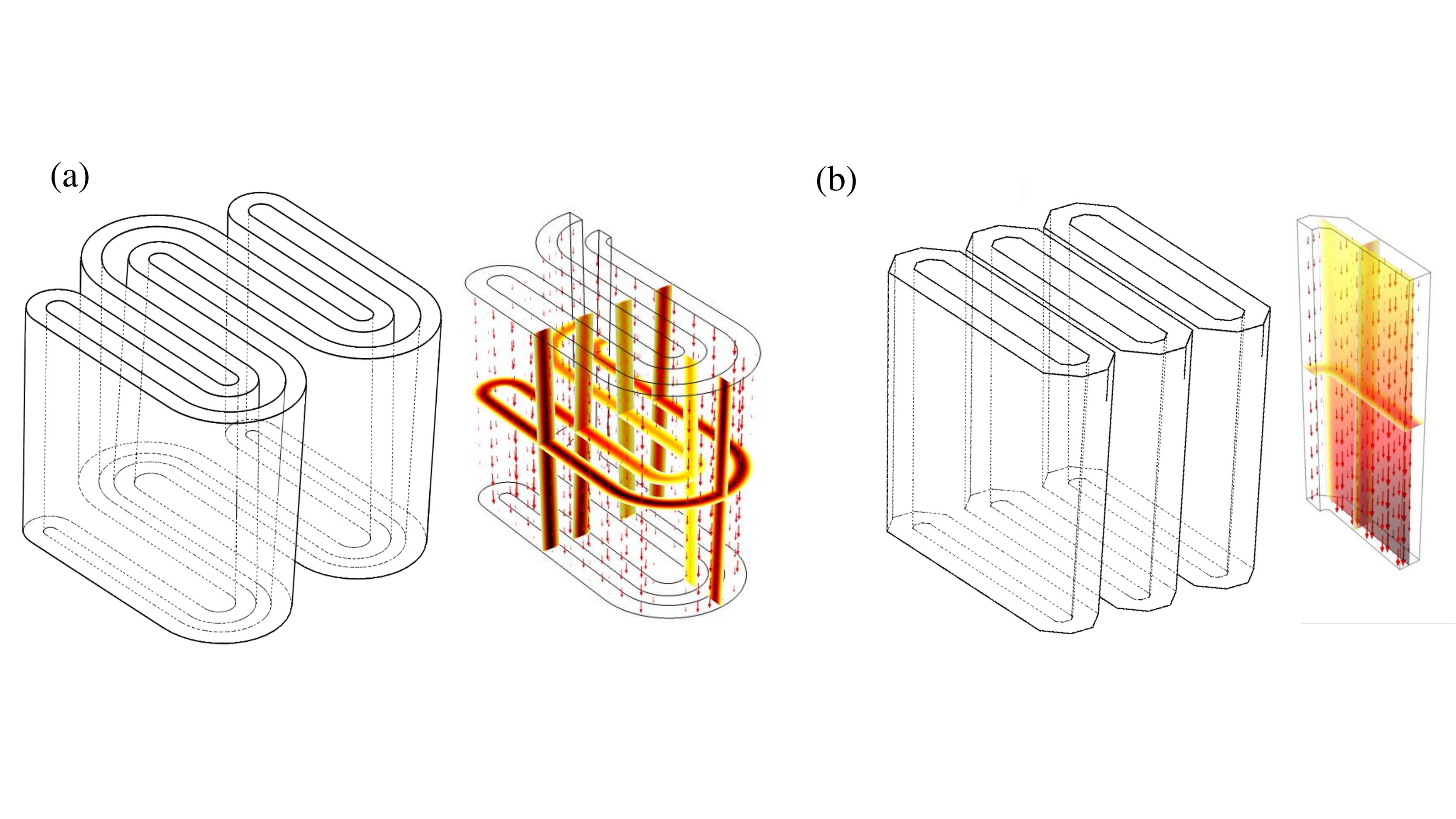}
\caption{\label{fig:eigenmode} The SolidWorks models and TM$_{010}$-like modes at $7.4$ GHz calculated using COMSOL-based FEAs for (a) a tunable ``W-Meander'', created following the procedure in Figure \ref{fig:W}, and (b) three stacked ``Racetracks''.  The FEA results show the $z$ component of electric field $E_z$ (color scale) and the $E$-field vectors (arrows). For both designs, the cone half-angle $\alpha=\tan^{-1}{0.025}=1.43^\circ$. 
Despite some noticeable $z-$ dependence in the field distributions, the form factors remain above $0.6$ over the entire frequency range for both cases.  The effective volume of these designs are plotted in Figure \ref{fig:EV}. }
\end{figure}

Another design of great interest is the ``Racetrack'' cavity, which is a squeezed variation of the cone cavity.  There are two advantages: (1) the planar surfaces might make it easier to achieve the required fabrication tolerances, and (2) the squeezed dimension saves space, leading to the possibility of stacking and combining multiple cavities with a summing network. (Figure \ref{fig:eigenmode}b) shows a stack of 3 racetracks.  As pointed out in Paper I, a summing network may already be necessary to read out a single cavity.  In Figure \ref{fig:EV}, the effective volumes for both a single (``Racetrack'') cavity and combination of three cavities (``3 Racetracks'') are shown. 

Figure \ref{fig:eigenmode} shows SolidWorks models of these two cavities and their TM$_{010}$ modes, which fill a large portion of the cavity volume.  The eigenmodes are shown for cavities tuned to the center frequency.  As pointed out earlier, better broadband performance is possible with global optimizations. Although the resulting $C_{010}$ from optimizing around the center frequency already seems satisfactory over the entire tuning range. 


There are several limitations to these results.  Symmetries have been assumed and invoked in the FEA.  To expedite the computations, the calculations for these cavities with very large volume ($V \gg \lambda^3$) are done with limited frequency resolution with no corrugations.  Frequency resolution is still too coarse to systematically study mode crossing.  Nevertheless, these results represent an existence proof for a TM$_{010}$-like mode in a volume-filling meander cavity.  The very high effective volume should strongly motivate future experimental investigations. Much more detailed results from a more modest, but fabrication-ready corrugated cavity are presented and discussed in the next section. 

\subsection{Corrugated Conic Shell-Cavity: A Case Study}

The corrugated conic shell-cavity depicted in Figure \ref{fig:cupholder} is an attractive concept both in terms of its ease of fabrication and the sizable effective volume (labeled ``Cup Holder'' in Figure \ref{fig:EV}).  This seems to be the first design to build before more elaborate structures are attempted. The performance of this cavity is therefore explored further.  
Also targeting a center frequency of $7.5$ GHz, the inter-cone distance $w$ is tunable from $18$ to $22$ mm by way of vertical motions of the inner cone. The diameter of the outer wall at the mid-point is $27.2$ cm; the cavity height is $24$ cm; and the cone half-angle $\alpha=\tan^{-1}{0.05}=2.87^\circ$.

\begin{figure}[b]
\centering 
\includegraphics[width=5.in]{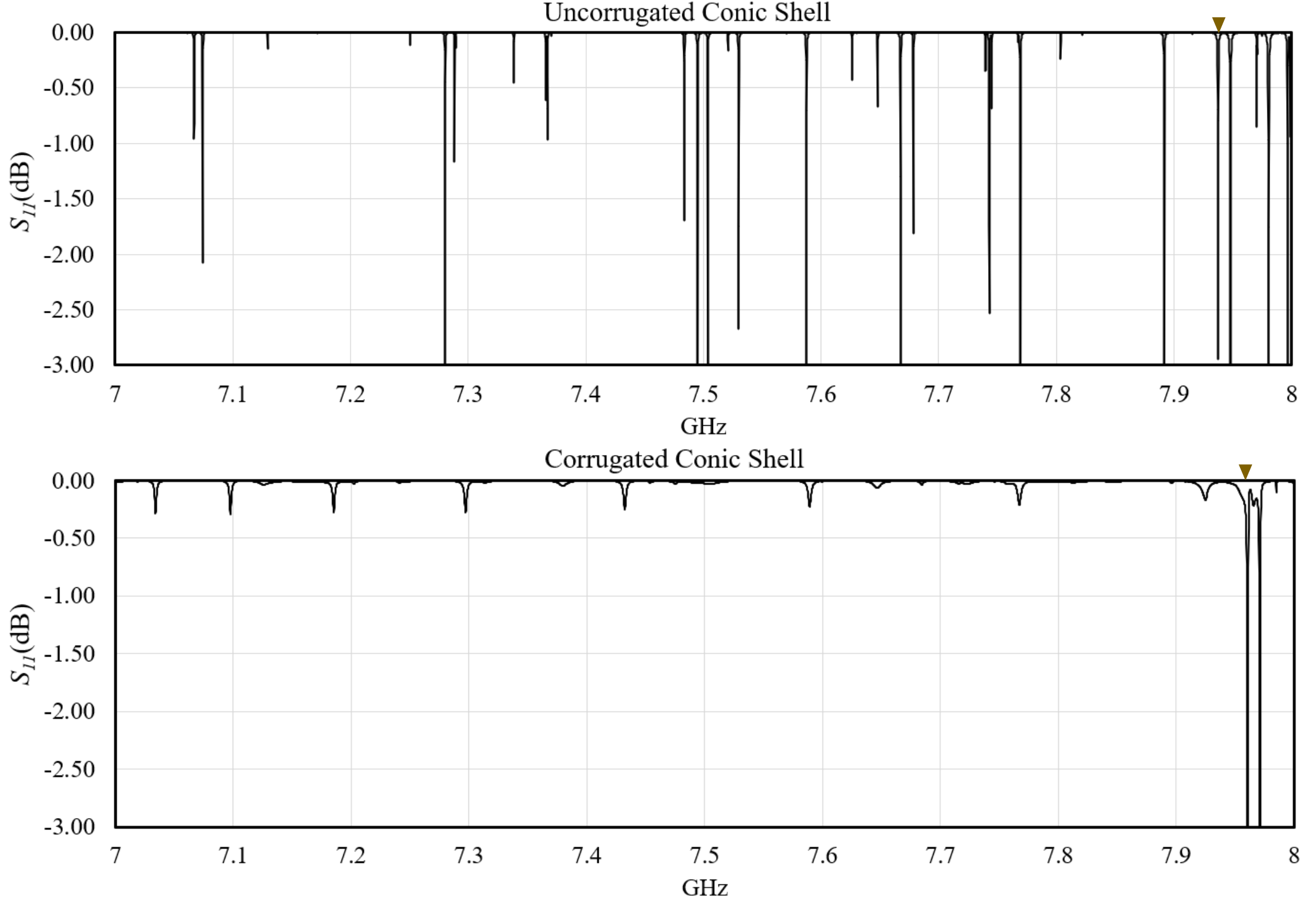}
\caption{\label{fig:S11} The $S_{11}$ parameters for uncorrugated ({\em Top}) and corrugated ({\em Bottom}) conic shell-cavities.  Both cavities are tuned to $w=18.9$ mm. The TM$_{010}$ mode appears at the marked locations ($7.9374$ GHz and $7.9602$ GHz). All of the TE modes and high order TM modes are located at higher frequencies and co-tuned with TM$_{010}$, inducing no mode crossings. 
The top panel shows many TEM resonances at frequencies below the TM$_{010}$ mode, creating potential problems from mode crossings. On the other hand, the TEM modes are noticeably suppressed in the corrugated cavity. }
\end{figure}

\begin{figure}[t]
\centering 
\includegraphics[width=5.4in]{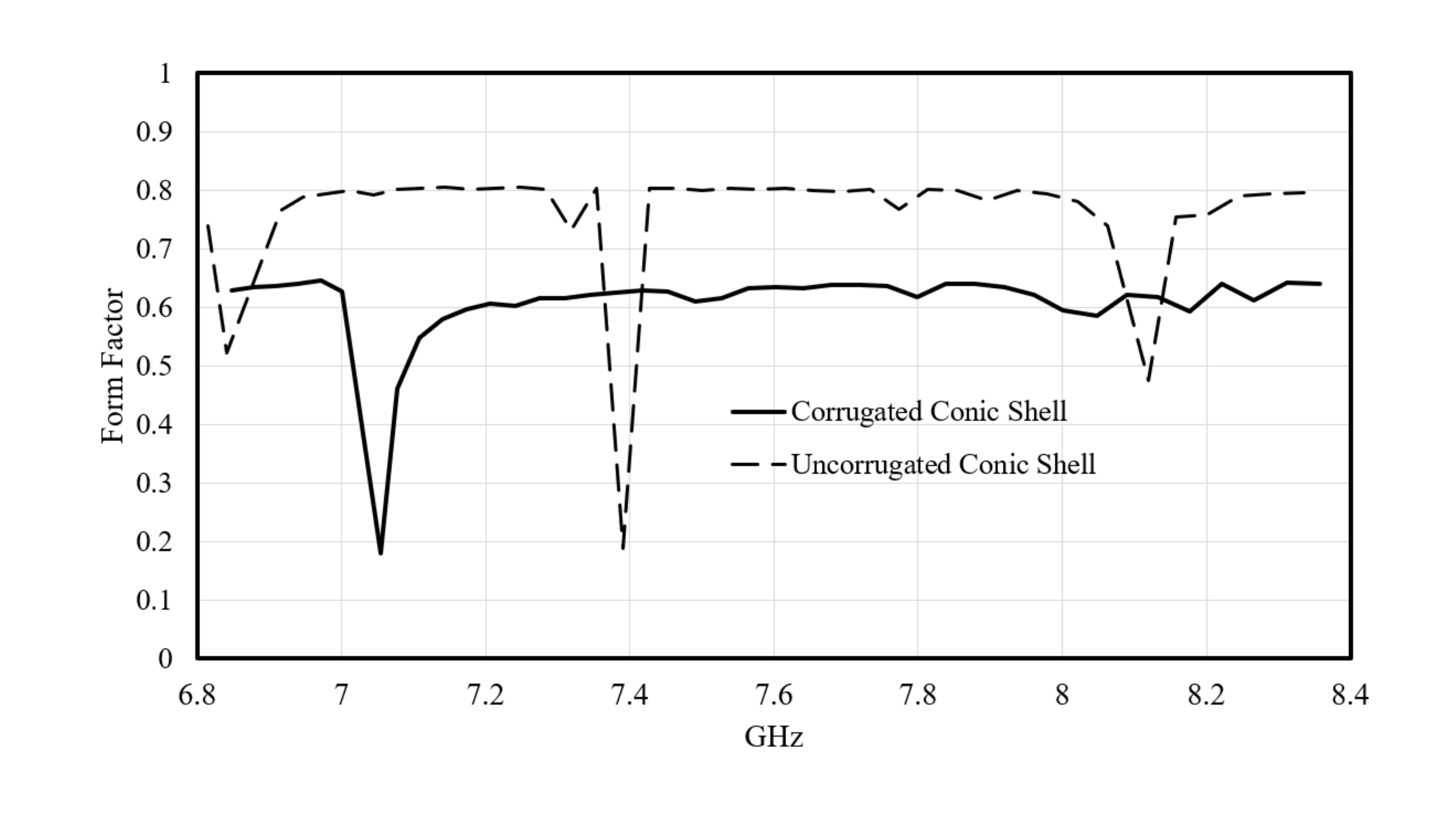}
\caption{\label{fig:cupformfactor} The form factor $C_{010}$ as a function of tuning frequency calculated from 3D FEA for two conic shell cavities (``Cup Holder'').  The solid line is for the corrugated cavity depicted in Figure \ref{fig:cupholder}.  Most of the $20\%$ tuning range is usable for axion searches.  The form factor of an uncorrugated conic shell cavity is plotted (dashed line) for comparison.  The corrugated cavity has the anticipated lower form factor \cite{Kuo_2020} but somewhat suppressed TEM mode-crossing.  
 }
\end{figure}

To accommodate tuning motions of the inner cone, gaps (Figure \ref{fig:cupholder}) of $4$ mm are added at the top and bottom (when $w$ is at the maximum).  This gap size corresponds to $0.09\lambda$ at the center frequency.  The regions outside the gaps are included in the FEA. No significant leakage is found with appropriate $\lambda/4$ corrugations. While symmetries have been invoked in the last section to obtain the TM$_{010}$ solutions for the W-Meander and Racetrack cavities, full-size 3D FEAs are performed for this cavity to survey all possible spurious resonances. Finally, a rectangular waveguide port is added for excitation to simulate the realistic set-up. The port is under-coupled by design to probe the intrinsic properties of the cavity. The resonances are identified in frequency sweeps of the $S_{11}$ parameter for each of the tuning configuration (Figure \ref{fig:S11}). The form factors at the TM$_{010}$ peak are calculated based on the $E$ field distributions excited by the waveguide coupling port. The high form factors obtained (Figure \ref{fig:cupformfactor}) directly verify that a local port can couple to a super-Compton resonant mode that fills the volume of the cavity.

A main goal of the detailed study in this section is to assess bandwidth losses due to mode-crossing using realistic 3D modeling.  
The effects of mode-crossing in axion haloscopes have been studied in detail \cite{stern_19,rapidis_19}.  The most egregious phenomenon that degrades the bandwidth is the so-called avoided crossing, which takes out a significant fraction ($\gg 1/Q$ for each crossing) of the originally usable bandwidth. The avoided frequency range is dictated by the amount of spurious coupling Hamiltonian between eigenmodes \cite{stern_19, Heiss_12}. 
Reference \cite{rapidis_19} reports that $15\%$ of the HAYSTAC bandwidth is significantly affected by avoided mode crossings.  

\begin{figure}[b]
\centering 
\includegraphics[width=5.in]{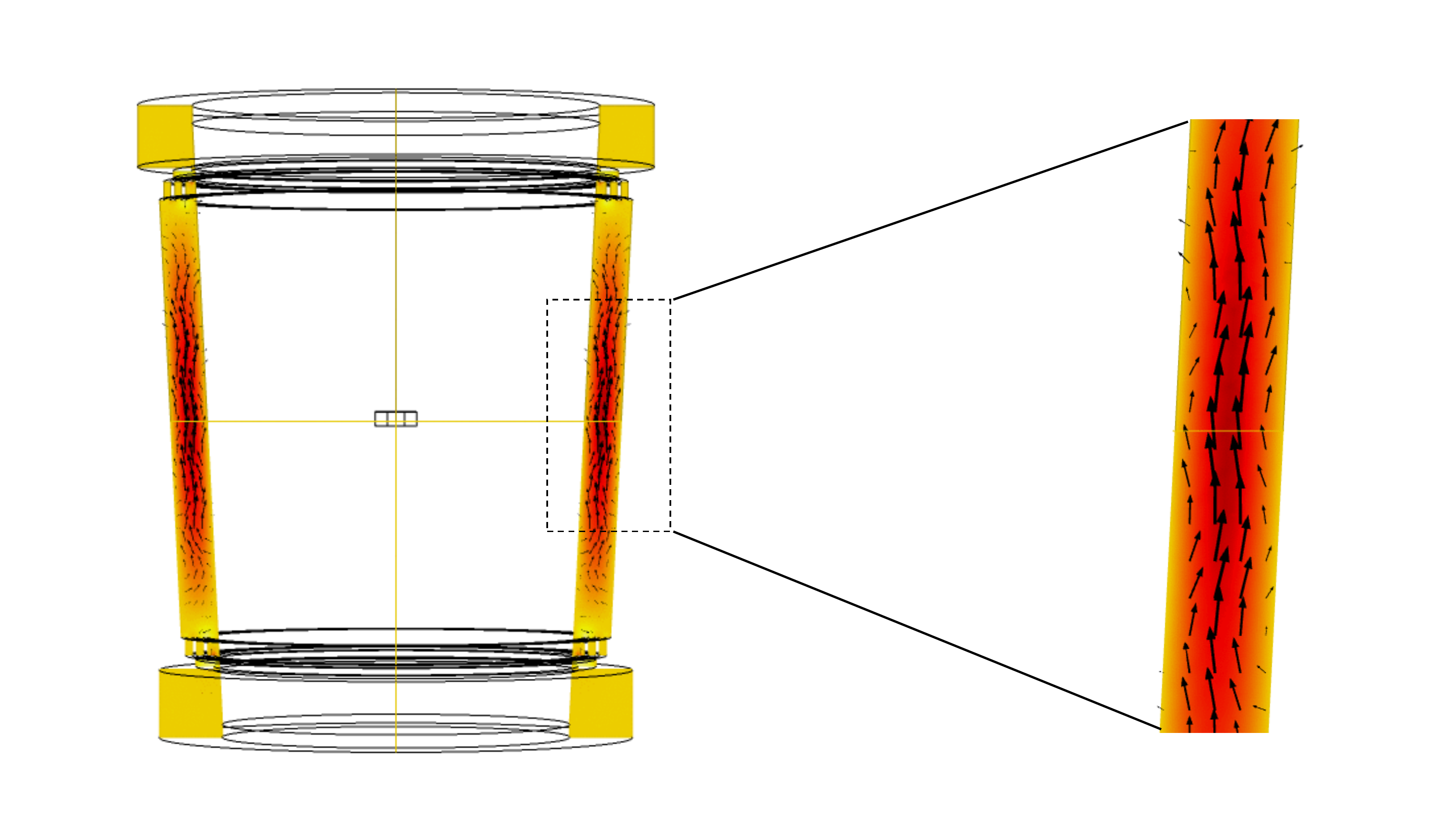}
\caption{\label{fig:wiggle} The electric field distribution of a TM$_{010}$-like mode in the corrugate cone cavity depicted in Figure \ref{fig:cupholder}.  The color scale is $E_z$, and the arrows represent the size and direction of ${\bf E}$ at each location.  The corrugations at the top and bottom confine the mode in the cavity.  No significant leakage is found through the gaps that enable mechanical tuning.  Minor radial undulations can be seen but the overall electric field is pointing in the $z$ direction. The amplitude of the radial undulation increases with cone half-angle $\alpha$. In this design, $\alpha=2.87^\circ$.  This eigenmode generally gives a form factor $>0.6$ for ${\bf E}\cdot {\bf B}$ coupling to an external solenoid field over a frequency range of $\sim 20\%$ (Figure \ref{fig:cupformfactor}). }
\end{figure}

Naively, avoid mode crossings could be much worse for large-volume over-moded cavities, rendering a large fraction of the bandwidth unusable.  This would have been a major roadblock for highly over-moded cavities.
To investigate, 40 numerical models of corrugated conic shell-cavities are calculated with approximately $0.5\%$ frequency spacing.  These exhaustive FEAs produce overall high ($\sim 0.61-0.63$) form factors across the tuning range (Figure \ref{fig:cupformfactor}).  The only exception is near $7.05$ GHz, where the form factor dips to $0.18$. This is indeed identified to be caused by an avoided mode crossing.  The form factor recovers to $0.46$ in the next frequency bin, and to $0.55$ in the following bin.  This survey already indicates that most of the frequency tuning range ($\gtrsim 95\%$) is usable for axion searches.

At frequencies away from such obvious mode crossing, TM$_{010}$-like eigenmode contains some undulations in the radial direction in addition to the expected dominant $E_z$ fields (Figure \ref{fig:wiggle}).  
After some experimenting, the cause of the undulations was identified to be neither the corrugations nor the waveguide port.
Rather, this can be understood to be low-level hybridization with otherwise hard-to-detect TEM modes. This phenomenon reduces the form factor slightly for a given cone half-angle but does not affect the main findings of this paper. 

This manageable level of mode mixing can be explained by two observations. 
First of all, the waveguide port naturally deselects $r$-polarized TEM modes, as pointed out in Paper I. Secondly, the results for the corrugated and uncorrugated cavities (Figure \ref{fig:S11} and Figure \ref{fig:cupformfactor}) suggest the strengths of the TEM modes are suppressed by the azimuthal corrugations. 
It should be noted that references \cite{rapidis_19} and \cite{stern_19} both find that the avoided crossing phenomena can be reproduced numerically if the models faithfully capture the features in the actual experimental set-ups.  
Avoided crossing in axion haloscopes is traced to spurious coupling between eigenmodes, often through coupling through the tuning mechanism.  For example, the spurious coupling that generates avoided crossings in cylindrical cavities is caused by the small gap at the mounting point of the tuning rod \cite{stern_19} which can be minimized but not completed avoided.  
One the other hand, in the corrugated conic shell design the cavity is defined by two truly disjoint pieces.  The corrugations and the sub-wavelength tuning gaps are already in the FEA model for Cup Holder cavity. No other features are anticipated to be present in the actual experiment. Therefore, the observed large tunable bandwidth should be representative of the actual set-up. Room-temperature prototyping will help confirm that. 

\begin{figure}[t]
\centering 
\includegraphics[width=5.6in]{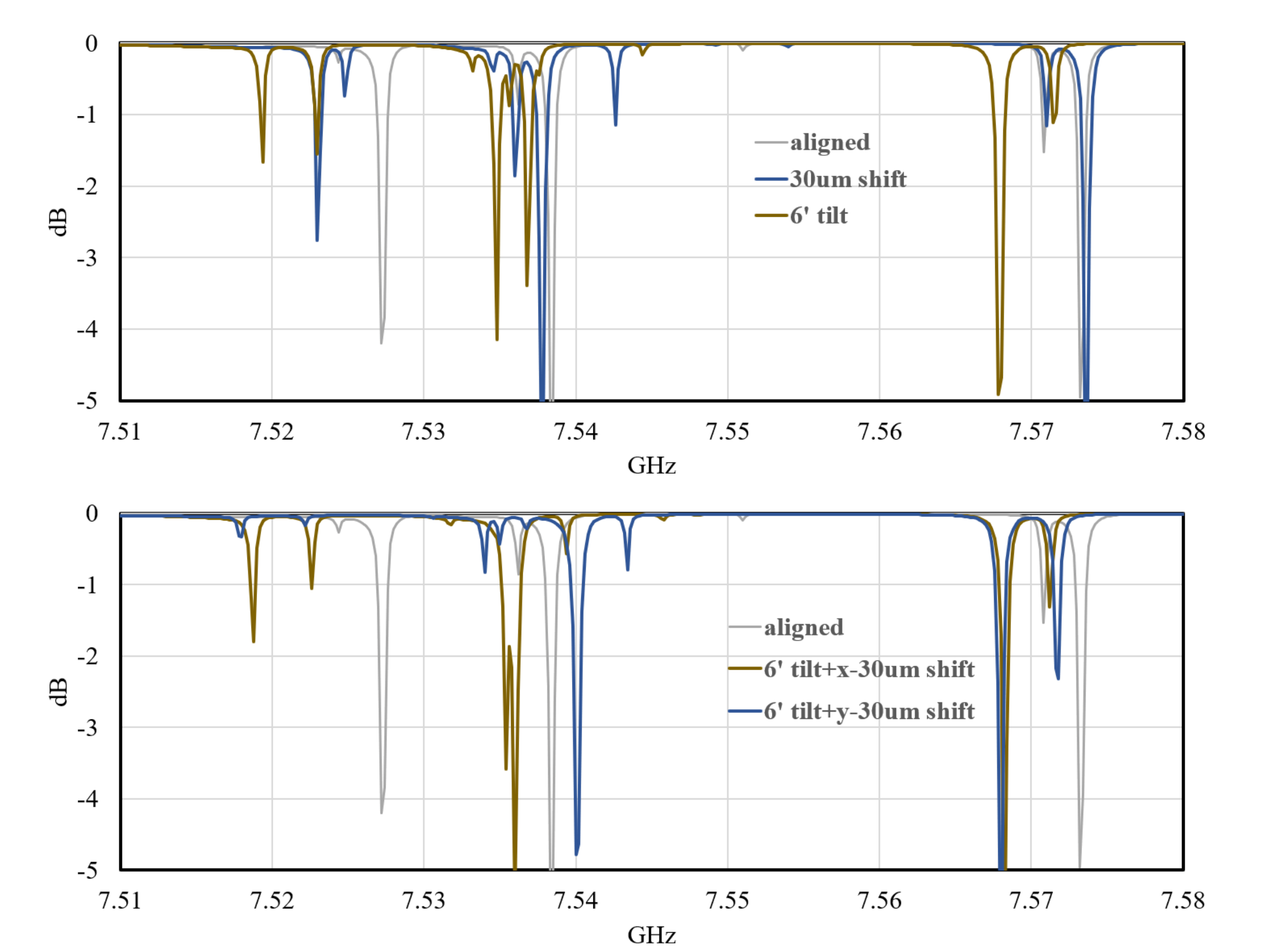}
\caption{\label{fig:alignment} The {\em in situ} microwave measurements can be used to align the inner cone with respect to the outer cone.  The two panels show the $S_{11}$ parameters of misaligned corrugated conic shell cavities overplotted on an ideally aligned cavity. The TM$_{010}$ resonance for axion searches is located at $7.5272$ GHz.  Tilting and displacing the inner cone beyond certain thresholds cause splitting and shifting of this resonance in complicated but predictable ways.  Misalignments can be identified in the software and adjusted mechanically in much the same way {\em in situ} stellar images are used for optical alignment and focusing in modern astronomical telescopes.  Based on these FEA, the required alignment tolerances are $\sim 10\; \mu$m and $\sim 3'$.
 }
\end{figure}

\section{Example of a Possible Experimental Realization}

Can we really make these complicated cavities? 
As discussed in Paper I, the distance $w$ between the inner and outer walls must be maintained to certain precision throughout the cavity shell. 
The relative placing of the two disjoint assemblies is also important for setting up the high-Q resonances. One of the six degrees of freedom (DoF) for placing the inner cone is associated with frequency-tuning. For circular cones the azimuthal DoF is degenerate (if the machining tolerance is met).  That leaves four more DoFs for alignments, which can just be $x,y$ linear displacements on both ends. The wealth of resonances provide good opportunities for precision alignments.  To demonstrate how that works, Figure \ref{fig:alignment} shows the $S_{11}$ parameter for a few misaligned Cup Holder cavities compared to a perfectly aligned one.  Symmetry-breaking leads to splitting of the TM$_{010}$ mode which can be identified in the computer software.  Misalignments can be corrected by making the eigenmodes with significant integrals of $E_z$ re-coalesce into a single resonance. Although {\em in situ} alignments necessitate cryogenic mechanisms, similarly demanding fine motion controls have been implemented successfully on orbital IR telescopes using the point spread functions measured from stars \cite{spitzer, akari}.

\begin{figure}[t]
\centering 
\includegraphics[width=6in]{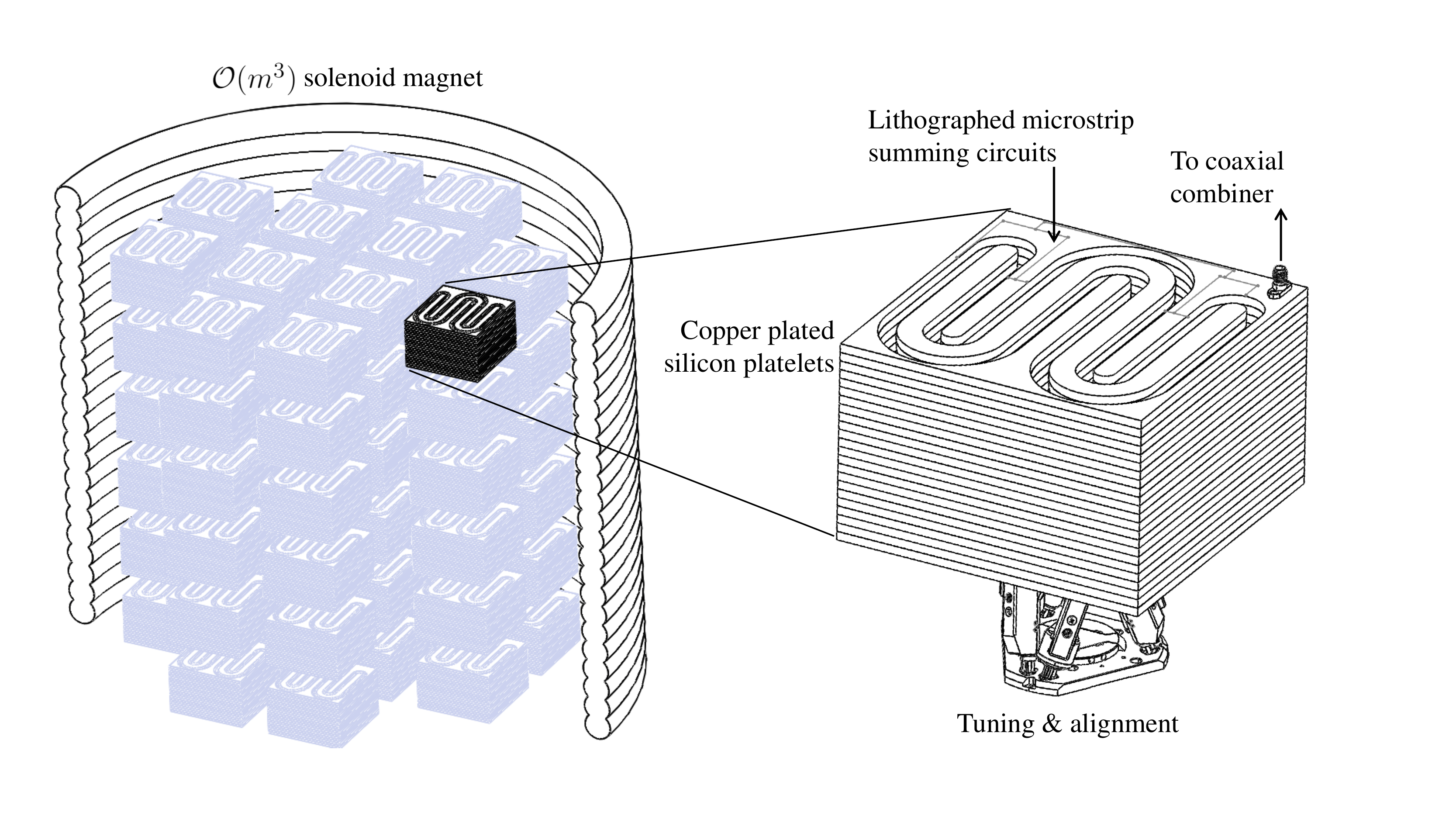}
\caption{\label{fig:axion20} A schematic diagram for the proposed cavity-based axion search at $\sim 20$ GHz. The required scan rate to reach the theoretical benchmark is achieved by instrumenting an array of brain cavities, each of which has the largest achievable volume.  To facilitate synchronous tuning and alignments, each cavity is equipped with own positioning mechanism such as the 6-DoF hexapod shown in the figure (reproduced with permission from Newport Inc.).  The fabrication of silicon platelet arrays is a well-established process in CMB instrumentation.  The overall architecture and scale of the experiment is similar to SuperCDMS.}
\end{figure}

The choice between ``large'' and ``many'' cavities is a trade-off between fabrication difficulties and alignment complexities.  The two arrangements in Figure \ref{fig:eigenmode}, the W-Meander and 3-Racetracks exemplify such trade-offs.  The fabrication of W-Meander is challenging, but it only requires 5 DoFs to align. The three Racetrack cavities are much easier to fabricate.  However,  the system has many disjoint pieces. To get a single high-$Q$ resonance, the experiment will require at least $3\times 6-1=17$ DoFs to align the inner shells to the outer shells.  It also requires an extra summing network to combine the signal \cite{jeong18,Kuo_2020}.  As another point of comparison, the new 7-rod (6 turning) HAYSTAC design \cite{Simanovskaia_20} (see also \cite{Stern_15}) will have $35$ DoFs for alignment and turn-synchronization. 

The effective volume of ADMX's cavity is around $8\times 10^{-2}\;m^3$ centered at a frequency $\sim 660$ MHz \cite{lyapustin}. At the higher end of the predicted dark matter axion mass in post-inflationary scenario ($>10$ GHz) \cite{diluzio_20}, the steep frequency scaling seems to put the DFSZ benchmark completely out of reach for these conventional cylindrical cavities.   
The large-volume shell cavity approach being advanced in this paper can conceivably be extended to above $10$ GHz. Such a high-frequency experiment centering at $20$ GHz may consist of $\sim 100$ ``large-volume'' cavities in an $\mathcal{O}(m^3)$ magnetic field (Figure \ref{fig:axion20}). Each cavity should be the largest realizable brain cavity for that frequency range.  Assuming the W-Meander cavity discussed in this paper can be practically implemented, each one would have a volume $\sim 0.5\times 10^{-3}\; m^3$ at 20 GHz (scaled from $\sim 10$ Liters at $7.5$ GHz). The total effective volume of the experiment is around $5\times 10^{-2}\;m^3$.
For comparison, a direct frequency scaling of ADMX cavity would lead to an effective volume of $3\times 10^{-6}\;m^3$. Although still a major endeavor, it is at least conceivable for the ``brain array'' approach to reach the theoretical axion benchmarks at $20$ GHz ($\sim 80\; \mu$eV).  

The fabrication methods for these high-frequency volume-filling cavities can borrow heavily from experiences and expertise in the CMB community. Like a CMB horn-plate \cite{britton_09}, the cm-wave brain cavity can be made by stacking DRIE (Deep Reactive Ion Etched) silicon layers known as the platelets.  The stacked assemblies are then plated with high conductivity metals that provide electrical continuity on the surfaces across the layers.  Each cavity has its own dedicated hexapod alignment/tuning mechanism. The top-most wafer can contain the lithographically defined superconduting microstrip summing network necessary to collect the signal from multiple ports in a meandering cavity. 
A schematic diagram of such set-up is shown on the right in Figure \ref{fig:axion20}.  
It is worth noting that the proposed architecture, fabrication techniques, and materials used are all very similar to those of a CMB detector module.  Over the next decade, several US facilities will produce hundreds of these CMB modules for the next-generation ground-based CMB observations \cite{cmbs4, so_19}. It is conceivable that a cm-wave axion search can leverage on that investment in infrastructure and R\&D. 
This rich technical heritage makes the brain array concept much more feasible.  

Another major advantage is that the axion-photon conversion happens in a solenoid magnetic field.  Unlike dipole or toroidal magnets, solenoid magnets require less custom engineering and can be available at a much lower cost.  One can also recognize the similarities in the overall set-up to the SuperCDMS (Super Cryogenic Dark Matter Search) experiment \cite{supercdms}. Fine mechanical positioning under extreme environmental conditions adds substantial complexities.  On the other hand, the fact that axion searches do not need to take place deep underground should simplify logistics of the project. 

Modern high speed digital electronics can greatly increase the readout and acquisition bandwidth  ({\em e.g.} \cite{umux}) and carry out the synchronization, alignment, and tuning in the computer software.  The detector end of the experiment is the same as that of a conventional haloscope.  Therefore, it works well in conjunction with novel quantum techniques that reduce readout noise, by either operating JPAs (Josephson Parametric Amplifiers) in a squeezed setup \cite{quantum1,quantum2}, or by using photon counting Qubits \cite{qubit2}.  However the frequency-matching between the cavities to a precision of $1/Q$ remains a major challenge. Individual dedicated hexapod tuning/alignment for each cavity should help in that regard. But it is possible that $\mathcal{O}(10^2)$ cavities can never to be tuned in unison. The scan rate would go as $V$ instead of $V^2$ in that case.

\section{Discussion and Conclusion}

The array of brains approach proposed in previous section is akin to taking the square root of a seemly insurmountable large quantity to make it smaller.  When facing such a large sensitivity gap, sometimes a brand new approach is needed.  Indeed, two genuinely new concepts have been proposed to search for axions at these high frequencies. 

\begin{itemize}
    \item {\it Dielectric Disks}  The first alternative proposal for high frequencies is to use a stack of high-index, low-loss dielectric disks to enhance the axion-photon conversion. This is the method adopted by the MADMAX experiment \cite{caldwell_17,madmax_17, madmax_19}.  The challenges for this concept have been openly discussed in a constructive way in the project's recent status report \cite{madmax_status_20}. Some of them include: the development and cost of a large-volume {\em dipole} magnet that provides a strong ($\sim 10$ T) $B$ field, the tiling of many smaller single-crystal dielectric pieces to form large ${\mathcal O}(m^2)$ disks, and the complexity associated with aligning and positioning (tuning) these (up to $80$) disks. It will become clearer whether the full-sensitivity proposal can be realized after its current prototyping phase ends. In some sense, the brain cavities and the MADMAX concept ({\em and} the original dish antenna haloscope \cite{Horns_2013}) all use axion-to-photon conversion that takes place in the presence of strong magnetic field parallel to electromagnetically active surfaces.  The quality factor $Q$ in cavities plays the role of the boost factor $\beta^2$ of a dielectric haloscope. The meandering surfaces in a brain cavity compactify the geometry and makes it possible to use a simpler magnetic system and to cool the entire system down to $100$ mK to reduce the thermal noise. 
    
    \item {\it Plasma Haloscope}  A more recent proposal is to match the plasma frequency of, {\em e.g.}, an array of evenly spaced metal wires to axion frequency \cite{mendon_19, plasma}.  There are a number of open questions regarding implementation, especially on signal readout and frequency tuning. A summing network as proposed in \cite{Kuo_2020} might be necessary to collect the signal from all parts of a plasma haloscope.  In this case, a significant challenge is that the summing tree needs to be fully integrated with the array of wires. Tuning by changing the spacing of the metal wires also seems hard. Reference \cite{plasma} points out other candidates for tunable plasma \cite{plasma2,plasma3}.  As mentioned in \cite{Simanovskaia}, experimental feasibility studies are being conducted on plasma haloscopes.  It remains to be seen whether any of these candidates can be set up, read out, and tuned on a sufficiently large scale.  
\end{itemize}

All three approaches, including the brain cavity array, come with substantial challenges.  As a community, more investigations and prototyping will be needed to determine the best approach to cover all the frequencies in this crucial axion mass range. 


\acknowledgments

The author acknowledges the support of a KIPAC Innovation Grant, and useful discussions with Blas Cabrera, Maxim Khlopov, and Marco Gorghetto.  The main idea of this paper was developed partly as a response to Alexander Millar's encouragement to find a way to tune the brain cavity proposed in Paper I.  

\nocite{*}

\bibliographystyle{JHEP}
\bibliography{ms.bib}









\end{document}